\documentstyle[12pt,psfig,aps]{revtex}
\newcommand{\gsim}{\,\raisebox{-.5ex}{$\stackrel{>}{\scriptstyle\sim}$}\,}
\newcommand{\be}{\begin{equation}}
\newcommand{\ee}{\end{equation}}
\newcommand{\bel}[1]{\be\label{#1}}
\newcommand{\ds}{\displaystyle}
\newcommand{\loo}{\,\raisebox{-.5ex}{$\stackrel{<}{\scriptstyle\sim}$}\,}

\begin{document}
\title{Coherent photon bremsstrahlung
and dynamics of heavy-ion collisions: comparison of different models}

\author{U. Eichmann$^\ast$, C. Ernst$^\ast$, L.M. Satarov$^\dagger$
and W. Greiner$^\ast$\\
 {${}^\ast$\,\normalsize \it Institut f\"{u}r Theoretische Physik,
 J. W. Goethe-Universit\"{a}t}\\
 {\normalsize \it D-60054 Frankfurt am Main, Germany}\\
${}^\dagger$\,{\it The Kurchatov~Institute,
123182~Moscow,~\mbox{Russia}}\\
}
\date{\today}
\maketitle

\begin{abstract}
Differential spectra of coherent photon bremsstrahlung in
relativistic heavy ion collisions are calculated within
various schematic models of the projectile-target
stopping.
Two versions of the degradation length model, based on
a phenomenological deceleration law, are considered. The
simple shock wave model is studied analytically.
The predictions of these models agree in the soft photon limit,
where the spectrum is determined only by the final velocity
distribution of charged particles. The results of these
models in the case of central Au+Au collisions at various bombarding energies 
are compared 
with the predictions of the microscopic transport model
UrQMD. It is shown that at the AGS energy the coherent photon bremsstrahlung
exceeds the photon yield from $\pi^0$-decays at
photon energies $\omega\loo 50$~MeV.
\end{abstract}

\section{Introduction} \label{sect_1}
Real photon bremsstrahlung emitted in the course of a heavy ion collision
may provide information about the underlying dynamics and the stopping law
of nuclear matter. 
To understand the sensitivity of photon spectra to
the dynamics of a heavy ion collision, it is instructive to
compare the predictions of models based on different
dynamical scenarios of the collision process. In earlier publications 
the coherent bremsstrahlung was studied in phenomenological
\cite{kapusta1,bjorken,thiel,lippert,dummi,eichmann,jeon,kapusta} as
well as microscopical \cite{koch} models. 
In the present
paper we concentrate on the comparative analysis of 
different models, which has not been done in the above mentioned works.

In section
\ref{sect_2} some general properties of bremsstrahlung spectra 
common to all models are obtained from basic
considerations about the nature of coherent bremsstrahlung.
section \ref{sect_3} introduces different 
deceleration models: a shock wave model 
and a degradation length model both for point-like and
extended nuclei. In addition we employ a microscopic
transport model which is not restricted to the initial
deceleration stage.
The comparison of these models is performed in section \ref{sect_4}.

A serious problem in observing the coherent photon
bremsstrahlung is the large background of
$\pi^0
\to\gamma\gamma$ 
decays. It is especially strong at
ultrarelativistic bombarding energies (AGS and higher). To overcome this
problem it was suggested \cite{dummi,jeon} 
to study experimentally
sufficiently soft photons (with energies less than several
MeV in the case of RHIC). According to our calculation 
the contribution
of the $\pi^0$ decays is indeed relatively small
at low photon energies. This problem will be studied
in more detail when we compare the models in section \ref{sect_4}.
Conclusions and discussion 
will be given in
section \ref{sect_5}.

\section{General Discussion}
\label{sect_2}
In this section 
some general model-independent properties of bremsstrahlung spectra are
discussed. 
According to the Larmor formula
the energy spectrum of the bremsstrahlung is proportional to the
squared absolute value of the Fourier transformed force $f(t)$
acting on a charge 
\cite{Jackson}, $dI/d\omega \sim
|f(\omega )|^2$.
From general properties of the Fourier transformation 
one is lead to a number of conclusions:\\
i)
At large photon
energies the radiated energy tends to zero
if the force is absolutely integrable. 
In contrast,
instantaneous interactions ($\delta$ forces) generate a flat energy
spectrum.  
The latter approximation is valid 
in the
so-called soft-photon limit ($\omega \to 0$), 
which in turn can be interpreted as a limit $\tau \to 0$ where $\tau$ is the
interaction time.\\
ii) From
the scaling behavior of
Fourier transforms
it may be shown that the bremsstrahlung drops
the faster in $\omega$ the longer the force acts on the charge 
(this is the analogue of the Heisenberg's uncertainty principle).
As a result, the 
energy spectra of bremsstrahlung photons 
become harder in the collision of smaller nuclei or at higher 
bombarding energies.
\\
iii)
If 
the force $f(t)$ 
is essentially nonzero in a finite time interval $|t|\le a$, 
which is a common assumption in all models
describing heavy-ion collisions far above the Coulomb barrier, 
the 
spectra of bremsstrahlung photons exhibit oscillations in $\omega$.
Indeed, 
any
continuous function on a finite interval
can be approximated with arbitrary accuracy by a
polynomial of a certain order $n$,
so that it is sufficient to consider only 
forces
of polynomial type. 
By subsequent integration by parts, 
it can be shown that the Fourier transform 
${f}(\omega)$
may be written as 
${f}(\omega) =
Ae^{i\omega a}+Be^{-i\omega a}$,
where  $A,B=\mp \sum\limits_{j=0}^{n}
\frac{i^{j+1}f^{(j)}(\pm a)}{\omega^{j+1}}$ and $A(B)$ corresponds to
the upper (lower) sign. 
Therefore 
the bremsstrahlung spectrum $dI/d\omega$
consists of a polynomial part
superimposed on an oscillating part. 
In this simple example the oscillating part has 
the frequency $\pi /a$. 
In real physical processes 
the frequency of the oscillations depends upon
details of the acting force. 

In a semiclassical approximation
the radiated energy per invariant phase-space element
is given by
\cite{ITC80}
\begin{equation}
\label{spektrum}
2 (2\pi)^3\frac{d^3I}{d^3k}=\,\mid j^*_\mu(k) j^\mu(k)\mid\; .
\end{equation}
where $I=\omega N$, $k=(\omega,\vec{k})$ 
is the photon 4-momentum and $j^\mu(k)$ is the
Fourier transform of the classical 4--current. 
By using charge conservation, $k_\mu j^\mu(k)=0$, and neglecting the
transverse components of $\vec{j}(k)$, which is a good approximation at
high bombarding energies, one has
\bel{scur}
|j^*_\mu (k) j^\mu (k)|=\left(\frac{k_\perp}{\omega}\right)^2|j_z(k)|^2\,.
\ee   
From Eqs. (\ref{spektrum}) and (\ref{scur}) it follows that 
\begin{equation}
\label{dido}
\frac{dI}{d\omega d\Omega}=\frac{k_\perp^2}{16\pi^3}|j_z(k)|^2\;.
\end{equation}
Let us assume that photons are produced in microscopic binary collisions
of point-like particles which instantaneously change their velocities in
each collision vertex. 
Then 
the explicit expression for $j^\mu(k)$ may be written as
\begin{equation} \label{gesstrom}
j^\mu(k)=i\sum_{i}\sum_j e_j\, \left(
\frac
{p^\mu_{ij}}{k\cdot p_{ij}}
-\frac{p^\mu_{i-1j}}{k\cdot p_{i-1j}}
\right) e^{ik\cdot x_{ij}}\; ,
\end{equation}
where indices $i$ and $j$ count vertices and particles,
respectively. The 4--vector $x_{ij}$ is the space--time position
of the $i$--th collision vertex of the $j$--th particle, $e_j$ is the
charge of the latter.
It is assumed that at $x=x_{ij}$ the
$j$-th particle 4-momentum jumps from $p_{i-1j}$ to $p_{ij}$\,.
Eq.~(\ref{gesstrom}) can easily be applied to inelastic particle collisions 
if one 
extends the definition of the charge $e_j \to e_{ij}$ and allow the charge to
change at a vertex. 
For example, if some charged particle is 
produced at the vertex $i$ 
then 
$e_{i-1j}\equiv 0$ and $e_{ij}\neq 0$.  

Assuming for simplicity that the particles 
move 
along the beam direction 
$z$ one immediately finds that the emission of soft photons is
suppressed if the particles are reaccelerated in the course of the reaction. 
The same phenomenon takes place if 
charges are produced and reabsorbed during a heavy--ion
collision.
By introducing a time cut 
one can decompose the current
into parts having fixed but opposite signs of the acceleration or
into parts corresponding to 
increasing and decreasing number of charged particles, respectively. 
It can be shown that 
the interference of the radiation 
from 
these two 
current components is destructive for soft photons. 

The soft photon limit of (\ref{spektrum}) 
is independent of the collision dynamics, but depends 
only
on the charged particles' velocities in the initial and final states
of the reaction.
In this case the sums over the
in- and outgoing particles can be replaced by integrals over 
the initial and final
velocity distributions.
Let us consider the soft-photon limit of (\ref{spektrum}) for
the case of symmetric central heavy-ion collisions,
assuming that the charges move along the beam axis.
For sharp initial and final velocity distributions
one finds 
\begin{equation} \label{splsk}
\lim_{\omega\to 0}\frac{dI}{d\omega d\Omega}
=\frac{4Z^2\alpha(\beta_i^2-\beta_f^2)^2\sin^2\vartheta\cos^2\vartheta}
{(2\pi)^2(1-\beta_i^2\cos^2\vartheta)^2(1-\beta_f^2\cos^2\vartheta)^2}\;
.
\end{equation}
Here 
$\beta_i$ ($\beta_f$) is the initial (final)
velocity of charged
particles, $\vartheta$ is the c.m. emission angle,
$Z$ is the total charge of each nucleus and
$\alpha=e^2/(4\pi \hbar c)$ is the QED fine structure constant.
The l.h.s. of 
Eq.~(\ref{splsk}) exhibits the typical behavior of quadrupole radiation,
which is a direct 
consequence of the symmetry of the considered system.

Using Eq.~(\ref{splsk}) it can be shown that 
the angle of maximum radiation equals
\[
\vartheta_{\rm max}={\rm arccos}\,
\sqrt{\frac{1-(u_++u_-)-i\sqrt{3}(u_+-u_-)}{2}}
\]
with
\[
u_{\pm}={}^3\!\sqrt{-\frac{\beta_i^2+\beta_f^2-\beta_i^2\beta_f^2}
{8\beta_i^2\beta_f^2}\pm \sqrt{\left(\frac{-4+2\beta_i^2+
2\beta_f^2-3\beta_i^2\beta_f^2}{4\beta_i^2\beta_f^2}\right)^3+\left(
\frac{\beta_i^2+\beta_f^2+\beta_i^2\beta_f^2} {8\beta_i^2\beta_f^2}
\right)^2}}\; .
\]
Clearly
for $\beta_i\gsim 0.9$ 
$\vartheta_{\rm max}$
depends quite weakly upon $\beta_f$ (if $\beta_f$ is not 
too close 
to $\beta_i$).
To a good accuracy 
it may be 
assumed that 
$\beta_f\simeq 0$,
which leads to 
\begin{equation}
\label{tmaxq}
\vartheta_{\rm max}=
{\rm arcctg}\,\gamma_i
\end{equation}
with $\gamma_i = (1-\beta_i^2)^{1/2}$. 
Substituting $\vartheta=\vartheta_{\rm max}$ into (\ref{splsk}) yields  
the following estimate for
the height of the maximum 
in the photon angular distribution
\begin{equation}
\label{maxhei}
\lim_{\omega\to 0}
\left.\frac{dI}{d\omega d\Omega}\right|_{\vartheta_{\rm max}}=
\frac{Z^2\alpha (\beta_i^2-\beta_f^2)^2(2-\beta_i^2)^2}
{(2\pi)^2(1-\beta_i^2)(2-\beta_i^2-\beta_f ^2)^2}
\end{equation}
If $\beta_f \simeq \beta_i$ the exact value of
$\vartheta_{\rm max}$ becomes smaller. In the case 
$\beta_f\simeq\beta_i\simeq 1$ one finds $\vartheta_{\rm max}\sim
1/(\sqrt{3}\gamma_i)$.
At large bombarding energies the photons emitted 
in a heavy-ion collision 
are strongly peaked in forward and backward directions 
and therefore practically do not 
overlap in phase space. 
In this case 
the interference term between the projectile and target photons 
becomes negligible
and the quadrupole spectrum may be well approximated by 
the sum of
the dipole spectra generated by each nucleus. For pure 
dipole radiation we find 
$\vartheta_{\rm max}={\rm arccos}\,\beta_i$ for $\beta_f\simeq 0$ 
and 
$\vartheta_{\rm max}\sim 1/(\sqrt{3}\gamma_i)$ for $\beta_i\simeq 
\beta_f\simeq 1$.
Striclty speaking,
however, 
using the approximation of 
dipole radiation in (symmetric) heavy-ion
collisions is not correct.

The same conclusion about the low sensitivity of the soft photon angular
distribution 
also holds 
for broad final velocity distributions, provided that
the mean final velocity 
is smaller than $\beta_i$.
In this case only the amplitude of the radiation is sensitive to 
the degree of stopping. 
This discussion shows that for a large range of nuclear stopping soft
bremsstrahlung alone can not serve as a good tool to measure 
the impact parameter \cite{bjorken}.
In the soft photon limit 
only the amplitude of the bremsstrahlung spectrum is
affected by both the involved charge and the final velocity distribution. 
The investigation of the soft photon spectrum (Eqs. (\ref{splsk})   
and
(\ref{maxhei})), however, shows that with increasing bombarding energy
the sensitivity of the amplitude on the final
velocity becomes extremely
weak for a large range of final velocities. 
For example, in the case of 
a Au+Au collision at RHIC energies ($y_i={\rm artanh}\,\beta_i\simeq 
5.4$) the amplitudes for the final c.m. rapidities 
$y_f=0$ and $y_f=3$ differ by less than $5\%$.

As is well known, scattering of particles 
exclusively along the beam direction excites only 
the $m=0$ multipole components of the radiation field. This implies that no
photons are emitted in and transverse to the beam direction. 
Of course, 
transverse scattering or the loss of the projectile-target symmetry 
of the 
current (e.g. in peripheral collisions \cite{vasak}) 
excite components with 
$m\neq 0$. As a consequence, nonvanishing radiation at $\vartheta\simeq 0$
and $\pi/2$ will appear 
\cite{eichmann}. 

Let us now consider the influence of the nuclear charge form factor on
the photon spectra. At large bombarding energies
particles move 
practically along the beam axis, and 
transverse sizes of the colliding nuclei remain almost unchanged 
during the most violent initial stage of a heavy-ion collision. 
It is assumed for simplicity that
the transverse form factor $F(k_\perp)$ can be factored, 
yielding
approximately
\begin{equation}
\label{formfak}
\frac{dI}{d\omega d\Omega}\approx |F(k_\perp)|^2
\left(\frac{dI}{d\omega
d\Omega}\right)_{0}
\end{equation}
where $(dI/d\omega d\Omega)_0$ denotes the spectrum obtained for 
slab-like nuclei.
This formula is well justified only 
for a certain class of charge density distributions 
(see e.g.  Sect.~\ref{schocksect}).
It becomes exact at very large bombarding energies, where $F(k_\perp)$
is given by the transverse Fourier transform of the nuclear thickness
function \cite{jeon}. 
The presence of the nuclear form factor 
$F(k_\perp)$ suppresses
the photon yield at transverse photon momenta 
$k_\perp \gsim 1/R$, where $R$ is the nuclear radius.
At fixed photon energy $\omega$, this suppression becomes stronger for
larger angles of emission 
with respect to the beam axis. 
As a result, with increasing bombarding energy 
structures of $dI/d\omega d\Omega$ are shifted to
smaller emission angles.

Another feature common to all models is that the 
compression of nuclear matter 
at intermediate stages of a heavy-ion reaction 
is correlated with  the characteristic time of the collision. 
In general, smaller collision times correspond to larger intermediate
densities and to smaller slopes of the spectra with respect to $\omega$.
Therefore, measurement of 
these slopes may provide some information about the
maximal compression of matter in a heavy-ion collision.

\section{Dynamical models of nuclear collisions}
\label{sect_3}

\subsection{Shock wave model}
\label{schocksect}
In the first model it is assumed that two plane shock waves 
moving with constant c.m. velocities,
$\pm \beta_{\rm sh}$, develop at the moment of the first contact of
the projectile and target nuclei. 
Within the shock wave model (SWM) it is postulated 
that the c.m. velocity of 
nuclear matter is
zero behind the shock fronts, i.e. at $|z|<\beta_{\rm sh}t$. It is
assumed that the initial nuclei are cylinders with 
effective radius $R$ and length $2R/\gamma_i$.
One can write the
relation $\varrho_0\gamma_i\frac{2R}{\gamma_i}\pi R^2=A$,
where \mbox{$\varrho_0=0.17\,{\rm fm}^{-3}$} is the normal
nuclear density and $A$ is the atomic number of the colliding
nuclei. In the case of Au nuclei,
$R\simeq 0.98A^{1/3}$~fm $\simeq 5.7$~fm.
This is slightly less than the geometrical value
$R_g\simeq 1.12A^{1/3}$~fm $\simeq\,6.5$~fm.

Before the collision, at
$t<0$, the longitudinal component of the electromagnetic
current can be written as
\bel{lcur}
j_z(t,\vec{r})=\frac{eZ}{A}\varrho_0\gamma_i\beta_i
\Theta (R-r_\perp)\,\left[\Theta (\beta_i t-z)\,\Theta (z-z_1(t))-
z\to -z\right]\,.
\ee
Here $\Theta (x)\equiv\frac{1}{2}(1+{\rm sgn}\,x)$\,, and 
$z_1(t)= -2R/\gamma_i+\beta_i\,t$\,, $\vec{r}_\perp (z)$ are
the transverse (longitudinal) coordinates of $\vec{r}$ with respect to the
beam axis. The first and the second term in the square brackets of
Eq.~(\ref{lcur}) correspond to the projectile and target nucleus,
respectively.  The shock fronts, appearing at $t=0$, reach the rear sides
of the colliding nuclei at
$t=\tau\equiv\frac{2R}{\gamma_i(\beta_i+\beta_{\rm sh})}$.
At $t>\tau$ compressed matter
starts to expand into the vacuum. We neglect the photon
production at this less violent stage of the reaction, assuming that
$j_z (t,\vec{r})$ vanishes fot $t>\tau$\,.
The explicit expression for $j_z$  at the intermediate stage, i.e. for
$0<t<\tau$, is given by Eq.~(\ref{lcur}) with the replacement
$\Theta (\beta_i t-z)\to\Theta (z+\beta_{\rm sh} t)$\,.

In accordance with Eq.~(\ref{spektrum}), the spectrum of bremsstrahlung
photons is determined by the Fourier transform of the 4--current
$j^{\mu} (k)$\,. 
After introducing the transverse density form factor of the initial
nuclei
\bel{trff}
F(k_\perp)=\frac{1}{\pi R^2}\int {\rm d}^2 r_\perp
e^{-i\vec{k}_\perp\vec{r}}
\Theta(R-r_\perp)=\frac{2J_1(k_\perp R)}{k_\perp R}
\ee
one arrives at the expression
\begin{equation}
j_z (k)
=Ze F(k_\perp) \beta_i \left[
\frac{1-e^{i\omega\tau\,(1+  \beta_{\rm sh}\cos{\theta})}}
{(\omega-\beta_i\,k_z)\,\omega\tau\,(1+  \beta_{\rm sh}\cos{\theta})}
+\frac{e^{i\omega\tau\,(1-  \beta_{\rm sh}\cos{\theta})}-1}
{(\omega+\beta_i\,k_z)\,\omega\tau\,(1-  \beta_{\rm sh}\cos{\theta})}
\,\right]\,.\label{ftlc}
\end{equation}
Equivalently, the Fourier transformed 
current can be calculated directly using 
Eq.~(\ref{gesstrom}), assuming 
homogeneously charged nuclear matter and replacing the sum by an integral.
In this calculation 
the space-time positions of the vertices $x_{ij}$ coincide with 
the positions of the
shock fronts.

Using Eqs.~(\ref{dido}), (\ref{ftlc}) yields the final
result (the respective formula 
given in Ref.~\cite{Bertulani} is incorrect)
\begin{eqnarray}
\frac{dI}{d\omega d\Omega}&=&\alpha\left(\frac{Z\beta_i\sin{\theta}}
{\pi\omega\tau}\right)^2 F^2(\omega\,\sin{\theta})\left\{
\frac{\ds \sin^2{\left[\omega\tau (1+\beta_{\rm sh}\cos{\theta})/2\right]}}
{\ds (1-\beta_i\cos{\theta})^2 (1+\beta_{\rm sh}\cos{\theta})^2}\right.
\nonumber\\
\label{shockspec}
&+&\left.
\frac{\ds \sin^2{\left[\omega\tau (1-\beta_{\rm sh}\cos{\theta})/2\right]}}
{\ds (1+\beta_i\cos{\theta})^2 (1-\beta_{\rm sh}\cos{\theta})^2}+
\frac{\ds \cos{(\omega\tau\beta_{\rm sh}\cos{\theta})}
\left[\cos{\omega\tau}-
\cos{(\omega\tau\beta_{\rm sh}\cos{\theta})}\right]}
{\ds (1-\beta_i^{\,2}\cos^2{\theta}) (1-\beta^{\,2}_{\rm
sh}\cos^2{\theta})} \right\}\label{finf}\,.  
\end{eqnarray} 
This
formula contains only one model parameter -- the shock wave velocity
$\beta_{\rm sh}$\,. 
It determines the duration of the collision and the
achieved compression of nuclear matter.  
Strong shocks whith large energy densities
of the stopped matter correspond to $\beta_{\rm sh}\ll 1$. 
On the other hand, from the baryon current conservation, 
it can be shown that in the limit $\beta_{\rm sh}\to 1$ the compression of
matter behind the shock front $\varrho/\varrho_0$ approaches its minimal
value $\sqrt{(1+\beta_i)/(1-\beta_i)}$. 
The soft photon limit clearly corresponds to $\omega\tau\ll 1$\,.
It may be verified 
that the soft photon limit is reached for $\omega$ or $\tau  
\to 0$.
In the lowest order the photon spectrum given
by Eq.~(\ref{formfak}) is obtained, where $(dI/d\omega d\Omega)_0$ is equal
to the r.h.s. of Eq.~(\ref{splsk}) with $\beta_f=0$\,.
As expected, in this limit the photon spectrum does not
contain any information about the parameters of the shock waves.

\subsection{Degradation length model}
In the second model we treat each nucleus as a set of $n$
 test particles with the charge $e (Z/n)$ 
moving along the beam axis. 
In the ground state of a nucleus 
the test particles are homogeneously distributed. 
It is assumed that these particles 
collide when their
trajectories intersect in the $t,z$ plane.
The stopping law
for the colliding
matter is  determined by the rule how the slopes of the
trajectories change at the collision vertices. 
In the model we have applied a simple stopping law \cite{hwa}
\begin{equation}
\label{deglen}
\frac{dp}{dz}=-\frac{p}{\Lambda}\;,
\end{equation}
where $z$ is the distance in nuclear matter covered by the test 
particles and $\Lambda$ is the so called degradation length.  

To implement 
(\ref{deglen}) into the degradation length model 
(DLM1) 
we postulate that 
the change of momentum of each test particle 
is determined by its number of collisions, i.e. by the number of 
trajectories crossed by the test particle during its path in nuclear matter. 
The computational procedure is based on 
the finite difference equation 
\begin{equation}
\label{deglen2}
p(z+\Delta z)=p(z)-p(z)\frac{\Delta z}{\Lambda}
\end{equation}
The l.h.s. of Eq.~(\ref{deglen2}) gives the momentum after the test particle 
collision, 
$\Delta z$ is a fixed value determined by the distance travelled by the 
test particles in
the ground state of the nucleus. 
The vertex positions and velocities entering Eq.~(\ref{gesstrom})
are provided by the
construction rule for the trajectories. In the limit $n\to \infty$ we again
treat a homogeneously charged matter and
unphysical oscillations in the photon spectra caused by a finite number
of test particles are washed out.
The bremsstrahlung spectrum is obtained by 
substituting (\ref{gesstrom}) into (\ref{spektrum}).

\begin{figure}
\centerline{\psfig{figure=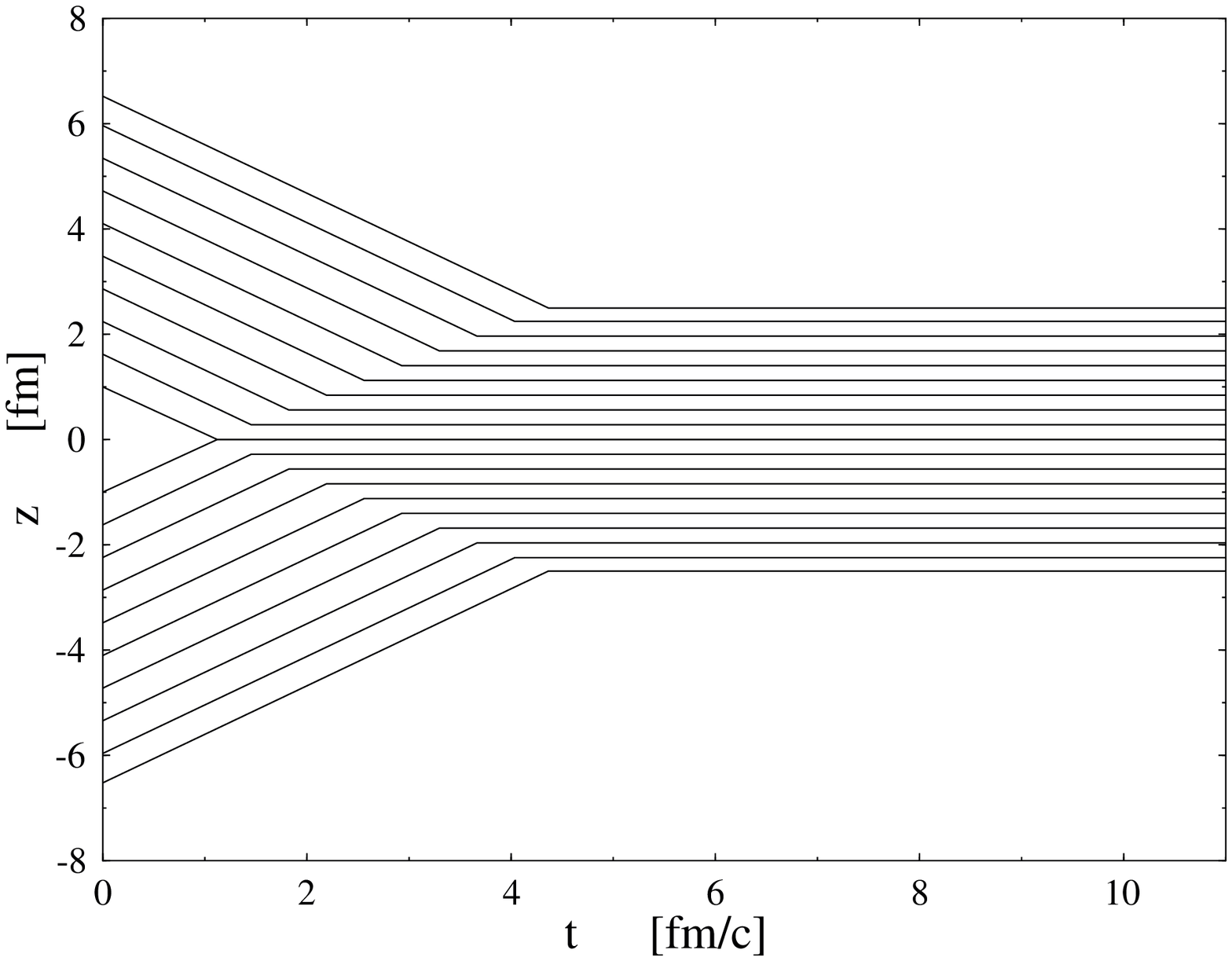,width=8cm}
        \psfig{figure=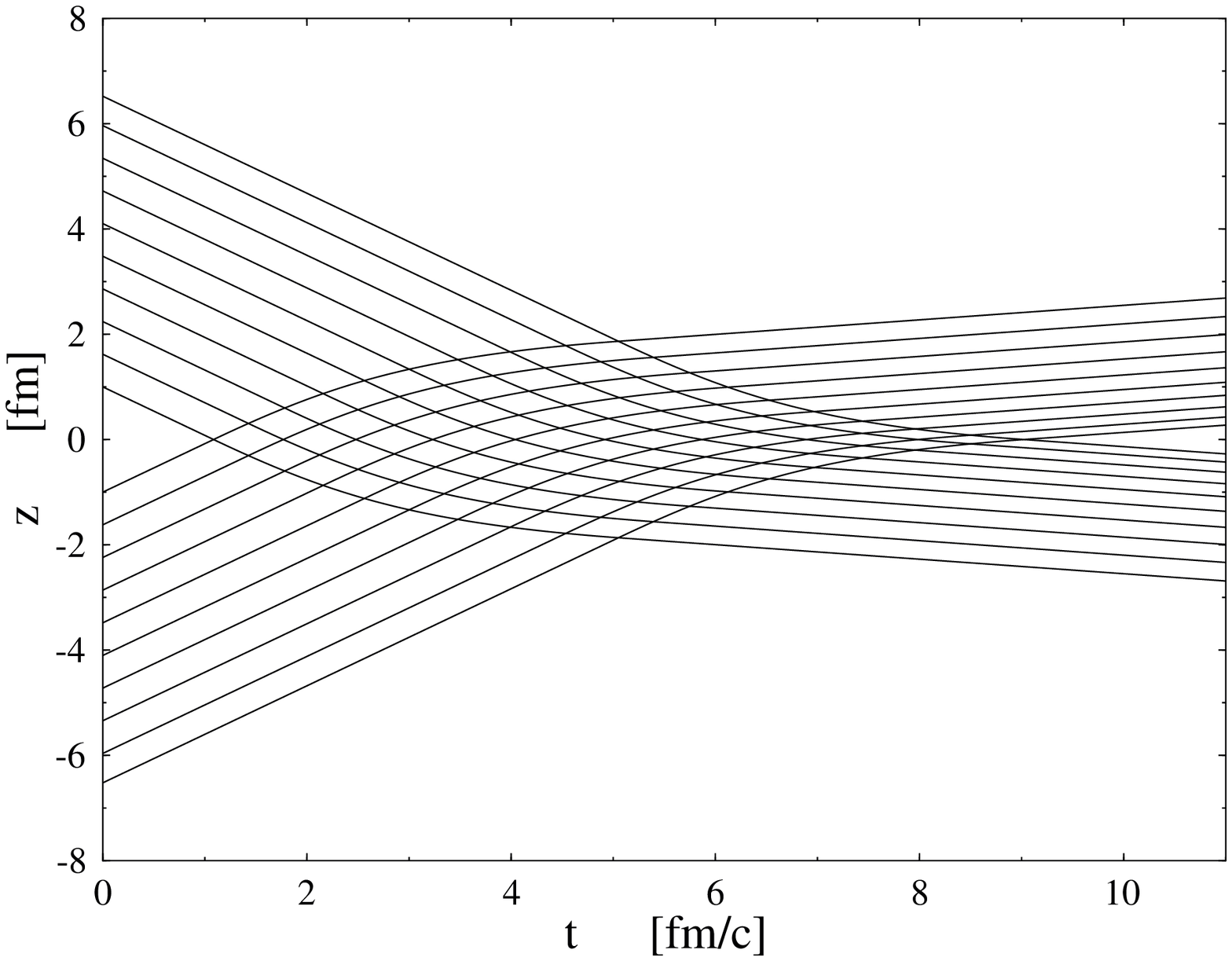,width=8cm}}
\caption{\label{traj}Trajectory representation of the shock wave model
(left) and the degradation length model (right). Both calculations represent
a central Au+Au collision at AGS energy (10.6 AGeV). }
\end{figure}

Figure~\ref{traj} 
shows the trajectory representations of the SWM 
and the DLM1. 
Dynamical differences between these two stopping scenarios
are clearly visible. One can see that in the SWM the 
stopping time 
is 
determined by $\beta_{\rm sh}$ and 
independent of the degree of stopping 
whereas in the
DLM1 this time 
depends on the degree of stopping.

In the extended degradation length model (DLM2) 
we generalize the DLM1 and consider
the colliding nuclei as homogeneously charged 
spheres represented by test particles. 
Binary collisions of test particles 
are treated in the same way
as above.
The final velocity of each test particle is again determined by the distance 
travelled in 
nuclear matter,  
but this distance now depends on the transverse 
separation from the nuclear center.
Therefore  spherical shapes 
of the colliding nuclei 
give rise to a broadening 
of the 
final
velocity distribution and
to a more complex structure of the current. 
On the contrary, the 
assumption of a cylindrical nuclear shape adopted by the SWM leads to a
sharp final velocity distribution ($\beta_f=0$). 
As discussed in Sect.~II, the broadening of the final velocity
distribution leads to almost the same shapes of soft photon spectra,
but reduces the amplitude of the photon yield.

\begin{figure}
\centerline{\psfig{figure=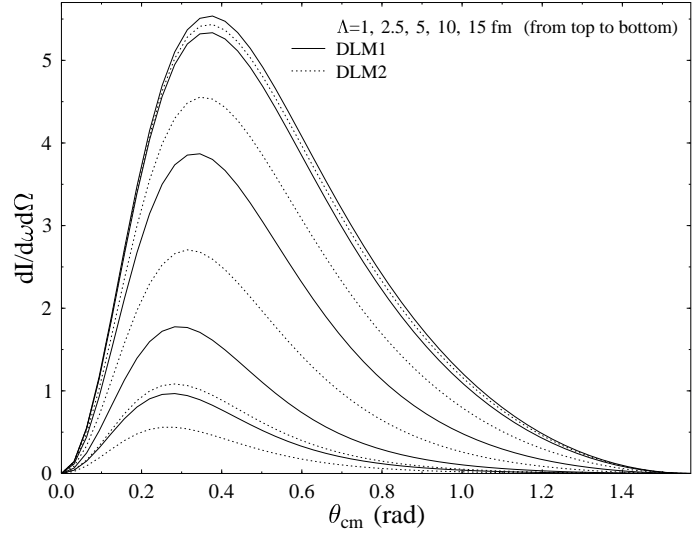,width=12cm}}
\caption{\label{lamver}
Angular distributions of soft photon bremsstrahlung ($\omega=1$~MeV) 
predicted by 
the degradation length model with and without nuclear form factor.
}   
\end{figure}

The influence of both the degree of transparency and the broadening of the
final velocity distribution due to the spherical geometry of the nuclei is 
demonstrated in Figure~\ref{lamver}. 
Since stopping is effectively reduced 
due to the spherical shape of the nuclei
the radiated energy is also reduced. 
The comparison of spectra predicted by DLM1 and DLM2 indeed 
confirms this conclusion. 
In accordance with the previous discussion
the sensitivity of the photon spectra to 
the final velocity distribution is low for small final velocities, i.e. in
the limit $\Lambda\to 0$. In
the case of weak stopping (large $\Lambda$) 
the maxima of the angular distribution 
shift 
to smaller angles. Calculations show that for
$\Lambda \loo 5$\,fm the angles of maximum 
radiation 
agree well with Eq.~(\ref{tmaxq}).  
Within the 
DLM1 the heights of the maxima are well explained by Eq.~(\ref{maxhei})
with the final velocity obtained from (\ref{deglen}). 
Alternatively 
Eq.~(\ref{maxhei}) 
may be used to estimate the mean final velocity in the DLM2.

\subsection{Transport model}
In the present paper we also consider the microscopic
transport model UrQMD \cite{Bass}.  In this approach the
whole history of the heavy ion collision enters
the photon spectrum. 
This allows to 
take into account 
the contributions from the late stages of the collision.

To demonstrate how the microscopic currents 
are treated 
in the
model it is instructive to rewrite Eq.~(\ref{gesstrom}) as
\begin{equation}
j^\mu(k)=i\sum_{i}\sum_j \hat e_j
\frac{p^\mu_{ij}}{k\cdot p_{ij}}
e^{ik\cdot x_{ij}}\; ,
\end{equation}
where the notation
\begin{equation}
\hat e_j = \left\{\begin{array}{ll}
e_j &\;\mbox{if $j$ corresponds to an incoming particle}\\
-e_j& \;\mbox{if $j$ corresponds to an outgoing particle}\end{array}\right.
\end{equation}
is introduced. The sum runs over all charged particles $j$
and their collision vertices $i$.
Such a procedure is applied to any kind of vertex (elastic
or inelastic collisions with arbitrary number of produced particles) as
well as for decays of unstable particles or strings.
Influence of mean--field potentials producing smooth
bending of particle trajectories is neglected.
Here it is assumed that hard binary collisions results in
much stronger decceleration of the particles.

An important background contribution to  
photon 
spectra comes from 
$\pi^0\to \gamma\gamma$ decays.  
In the UrQMD model 
the contribution of these decays
can be calculated
directly. 
In this calculation it was assumed that
$\pi^0$ mesons decay isotropically in their rest frame.

To reduce the statistical fluctuations of the calculated
spectrum the latter was averaged 
over a large
number of events.

\section{Comparison of photon spectra}
\label{sect_4}
In this section we compare the model predictions in the case of 
central Au+Au collisions at various bombarding energies
from GSI to RHIC. Since photon radiation in a symmetric
heavy--ion collision
is symmetric in the c.m. frame with respect to $\vartheta=\pi/2$,
the results are shown only for
$0\leq \vartheta\leq\pi/2$.

\begin{figure}
\centerline{\psfig{figure=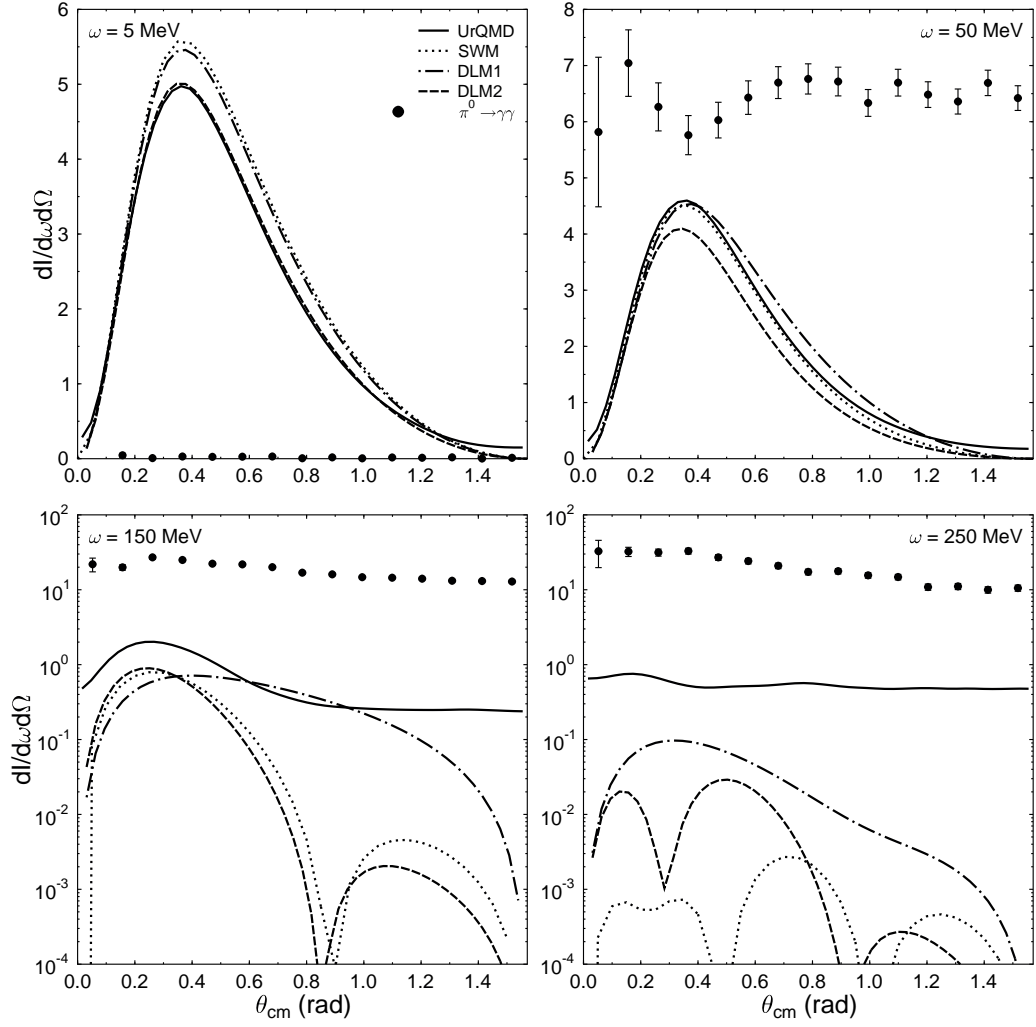,width=15cm}}
\caption{\label{angcomp}Angular distribution of bremsstrahlung spectra
predicted 
by different models 
for a central $10.6$~AGeV Au+Au collision 
at the photon energies $\omega=5,~50,~150$ and
$250$~MeV. The dots show the contribution of $\pi^0\to \gamma\gamma$
decays.}
\end{figure}

Figure~\ref{angcomp} shows the angular distribution of bremsstrahlung 
photons at the AGS bombarding energy 10.6 AGeV.
Four different photon energies are considered.
The parameters of the models are $y_{\rm sh}={\rm artanh}\,\beta_{\rm sh}=1$ 
(SWM) and
$\Lambda=2$\,fm (DLM). In the considered reaction
such a 
degradation length corresponds to rather strong stopping.

At $\omega=5$~MeV 
the SWM and the DLM1 
overestimate the bremsstrahlung yield as compared to the UrQMD
calculation. 
The inclusion of a realistic nuclear shape, however,  
results in a good 
agreement  of the DLM2 
with the UrQMD model. 
This agreement favours a strong 
stopping at intermediate stages of a central Au+Au collision at AGS energy.
Note that the degradation length $\Lambda$
entering the model calculation 
$\Lambda=2$~fm 
corresponds to $\Lambda^\prime\simeq 5.2$~fm in
the laboratory frame.  
The latter is close to the 
value obtained by fitting $pA$ data 
\cite{csernai}. 

All simple models incorporating only 
the stopping phase of a heavy ion                                        
collision underestimate the radiation in forward 
($\vartheta\simeq 0$) and 
transverse ($\vartheta\simeq \pi/2$) direction to the 
beam. This is  especially visible 
at larger $\omega$'s. 
As will be shown below, the additional contributions 
originate mainly from the late 
stages of the heavy-ion collision. 
Indeed, contributions from 
the expansion phase of the collision, and bremsstrahlung produced in 
rescatterings with newly produced particles
are disregarded in the SWM and DLM. 
It should be noted that at the considered energy the 
transverse scattering of particles 
(see 
Ref.~\cite{eichmann}) gives only small
contributions  
and cannot explain the above-mentioned discrepancy. 

We find that up to $\omega=50$~MeV all 
models predict roughly the same angular distributions. 
In this case photon spectra are
dominated by the quadrupolar radiation  
produced in the course of mutual deceleration of the colliding nuclei. 
As discussed above, 
bremsstrahlung photons with energies less than or of the order
$1/\tau$, where  $\tau$ is the
collision time, cannot resolve the detailed structure of the current. 
On the other hand,  
the simple models predict a more rapid decrease with $\omega$ 
than
the UrQMD model. 

Within the SWM and the DLM2 
the decrease of the photon yield at $\omega \gg 1/\tau$ is  
determined mainly 
by the nuclear form factor. 
The shift of maximum radiation  to smaller
angles (see Sect.~\ref{sect_2}) 
is clearly visible already at $\omega = 150$~MeV. 
As compared to the DLM1,  
the inclusion of the form factor in the
DLM2 leads to the appearance of additional maxima.
Note that predictions 
of 
the SWM and the DLM2 are rather close, although these models are based on
very different dynamical scenarios.  
The SWM predicts additional maxima 
with comparable heights and locations.
It is interesting that
these maxima appear in this model 
for any $\beta_{\rm sh}\neq 0$ even for slab-like nuclei. 

At $\omega = 250$~MeV angular distributions calculated within 
the simple stopping models
and the UrQMD model differ considerably both in shapes and 
amplitudes of the photon spectra. 
At this energy additional maxima appear also in the UrQMD model. 

As noted above,  
the spectra predicted by  the SWM and the
DLM decrease faster with respect to $\omega$ 
as compared to the spectrum calculated within the UrQMD
model. 
Special analysis shows that 
such a behaviour 
does not originate 
solely from different collision times included in these models.
Indeed,
at the AGS energy the spectrum predicted by the SWM
decreases faster with $\omega$ 
than the DLM2. 
On the other hand,
the characteristic collision time of the DLM2 
($\tau \sim 8$\,fm/c) is more than twice as large
as compared to the SWM
($\tau\sim 3.5$\,fm/c). A rough estimate shows that the
collision time of the UrQMD model is of the order of $6$\,fm/c.

The freedom in the choice of model parameters is limited
by the assumed stopping law.
For instance, it is not possible to improve the agreement 
between the DLM2 and the UrQMD model 
at higher photon energies 
(by chosing a smaller $\Lambda$) 
without losing agreement for soft photons. On the other hand, 
in addition to the
nuclear form factor, only the magnitude of the 
collision time
determines the slopes of the photon spectra within the SWM and DLM2.

\begin{figure}[hbt]
\vspace*{-1cm}
\centerline{\hbox{\psfig{figure=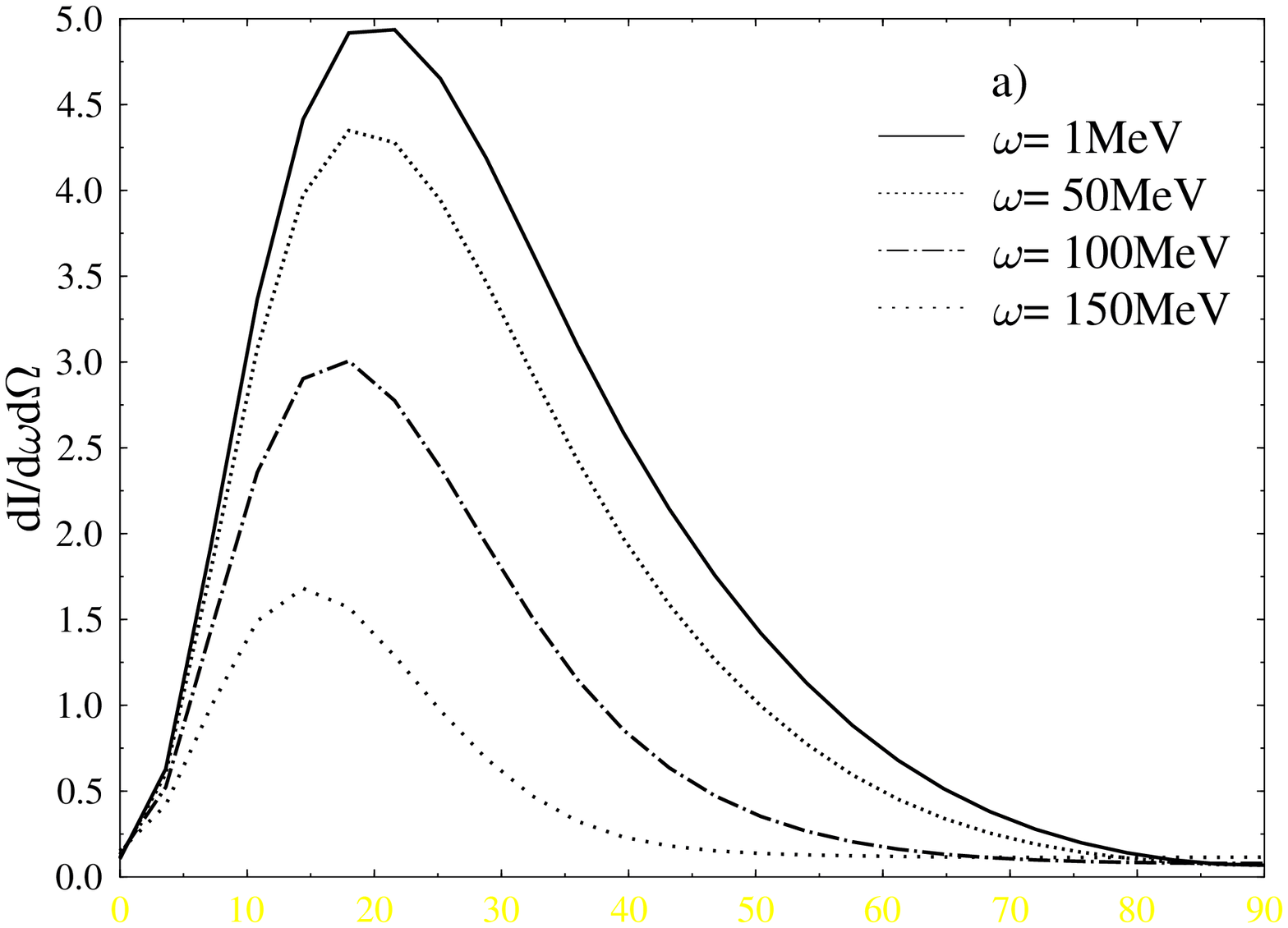,width=9cm}}}
\vspace*{-2.08cm}
\centerline{\hbox{\psfig{figure=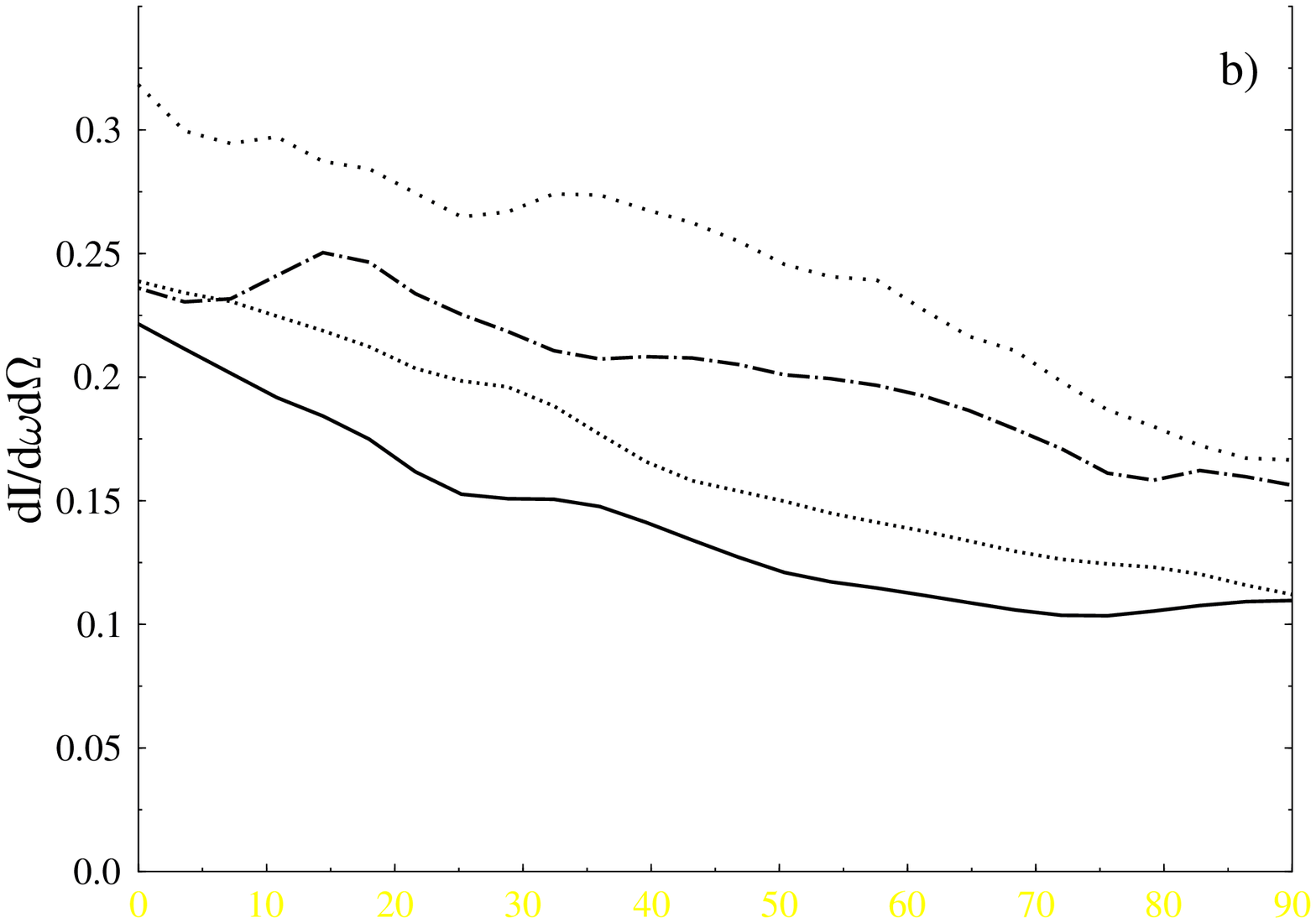,width=9cm}}}
\vspace*{-2.08cm}
\centerline{\hbox{\psfig{figure=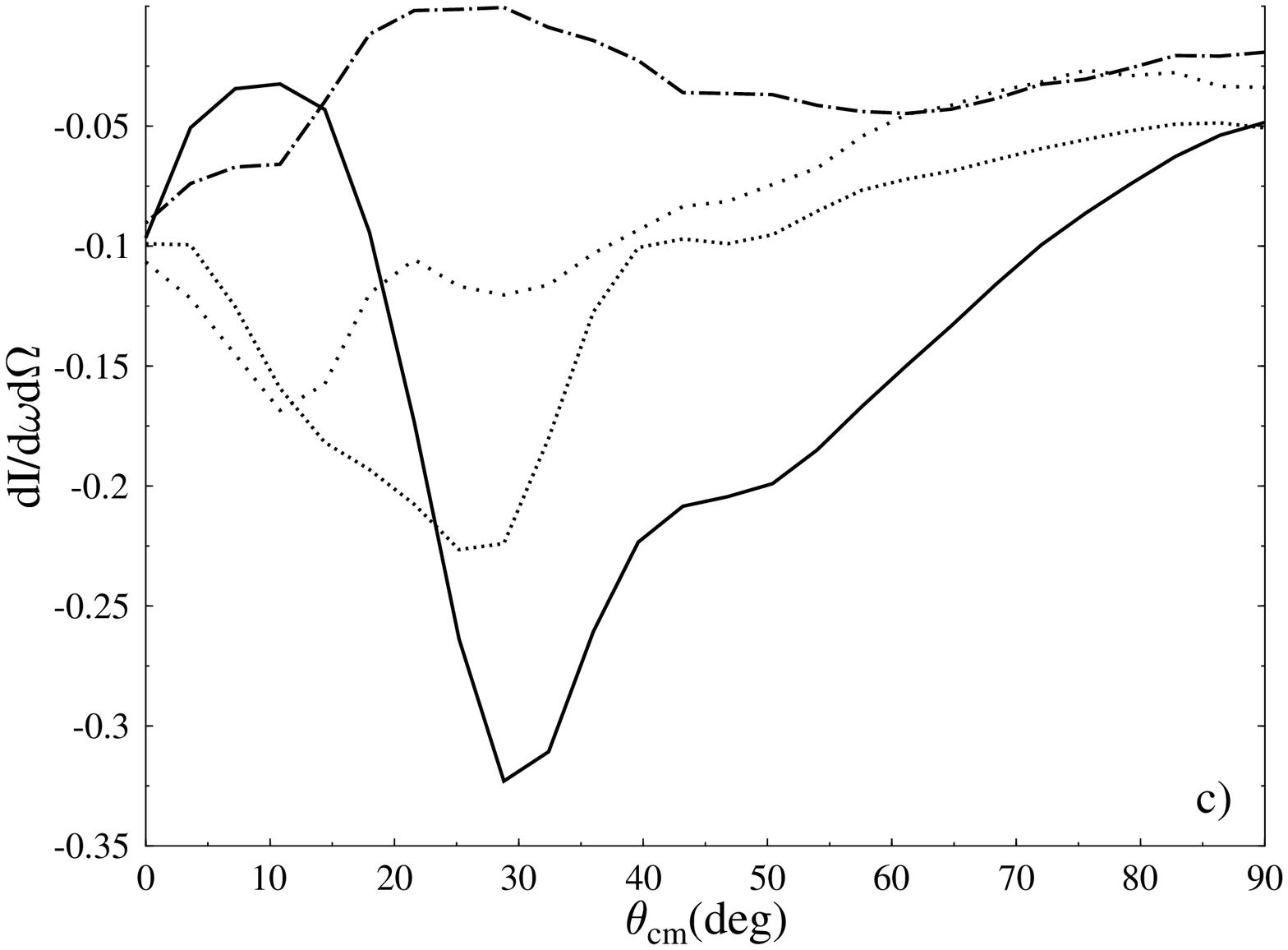,width=9cm}}}
\caption{\label{decomp}
Contributions to the bremsstrahlung spectra from different stages of a central
Au+Au collision at AGS energy:
a)~deceleration
stage ($t<t_{\rm cut}$), b)~expansion stage ($t>t_{\rm cut}$), 
c)~interference of a) and b).}
\vspace*{-0.5cm}
\end{figure}

To clarify the situation, a more detailed study of the photon spectra has
been performed within the UrQMD model. 
Figure~\ref{decomp} shows the different components of the bremsstrahlung 
spectra 
generated
in UrQMD events. 
A time cut at $t_{\rm cut} \simeq 6$~fm 
has been introduced to 
separate the contributions from 
the early 
stopping phase and the late stages of
the collision. 
This time cut is 
close to 
the passage time of the colliding nuclei
$t\sim 2R/(\gamma_i\beta_i)$ where 
$R\sim 7$~fm is the geometric
radius of the gold nucleus.
The contribution of the stopping
phase ($t<t_{\rm cut})$ has the familiar quadrupolar shape
and is well understood within the SWM and DLM.
Bremsstrahlung from the late stages is nearly isotropic and 
is smaller for 
soft photons. 
This is understandable since the 
charge-current 
varies smoothly with time at these stages of the collision. 
The characteristics of the bremsstrahlung component from
$t>t_{\rm cut}$ do not contradict the conclusion 
made in Ref.~\cite{larissa} that local
equilibration is achieved in 10.6 AGeV 
central Au+Au collisions at
$t>$10~fm/c within a volume of $(5~{\rm fm})^3$.
Apparently, deviations from angular isotropy can be 
explained  by a nonspherical shape of the dense central region
elongated in the beam direction~\cite{larissa}.

The interference between the bremsstrahlung photons from the
two considered stages 
indicates to which extent it is justified to study the radiation
from the different stages separately. 
From the comparison of Figs.~\ref{decomp}a and \ref{decomp}c one can see 
that the relative contribution of this
interference is of the order of $5-10\%$. 
The noticeable negative interference at small photon energies 
probably follows from
a reacceleration of 
charges at late stages of the collision.
It may serve as a
signal of the final expansion of the compressed matter.  
Suppression of 
soft photons due  
to reacceleration of charges at late stages 
was found also in the 
Landau fluid-dynamical model \cite{kapusta}.
Due to the large interference 
the bremsstrahlung
models considering only the stopping phase of a heavy-ion
collision may underestimate the stopping power of nuclear matter 
if such models are used to 
fit the observed soft photon spectra.

\begin{figure}
\centerline{\psfig{figure=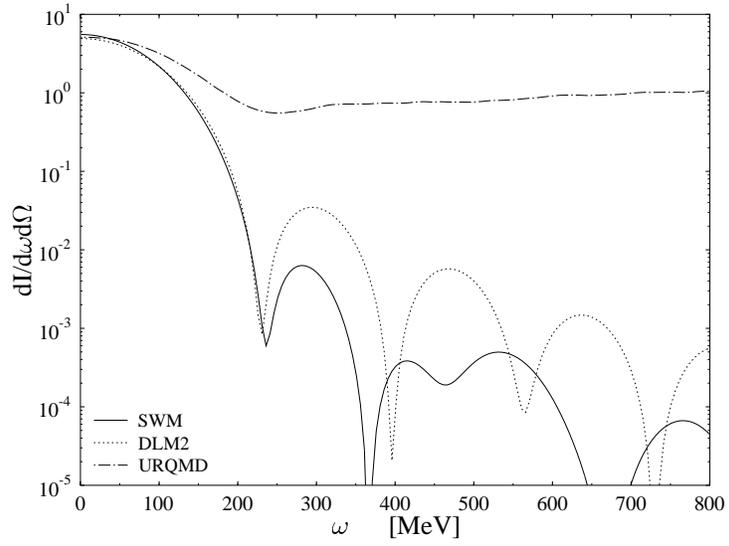,width=12cm}}
\caption{\label{osci}
Energy spectra of bremsstrahlung photons in a 10.6~AGeV central Au+Au
collision. }
\end{figure}

Figure~\ref{osci} represents the energy spectra of photons predicted with
the SWM, DLM2 and UrQMD model at the angle of maximum radiation given by
Eq.~(\ref{tmaxq}). 
One can see a qualitatively different behaviour of the spectra predicted by 
simple
models and the UrQMD model 
at large $\omega$'s. In the considered case of the AGS
energy the SWM and DLM2 noticeably underestimate the UrQMD predictions at
$\omega \gsim 150$~MeV. Moreover, the UrQMD spectrum slightly increases with
energy at $\omega \gsim 200$~MeV. Apparently such a behaviour also 
originates
from the late stages of the reaction.

As a result 
one may conclude that non-soft domains of the photon spectra can not
be studied realistically without an 
explicit treatment of the late stages of a
heavy-ion collision. On the other hand, the presence of the
$\pi^0$-background  highly complicates the measurements of bremsstrahlung
spectra at large photon energies. 

\begin{figure}
\centerline{\psfig{figure=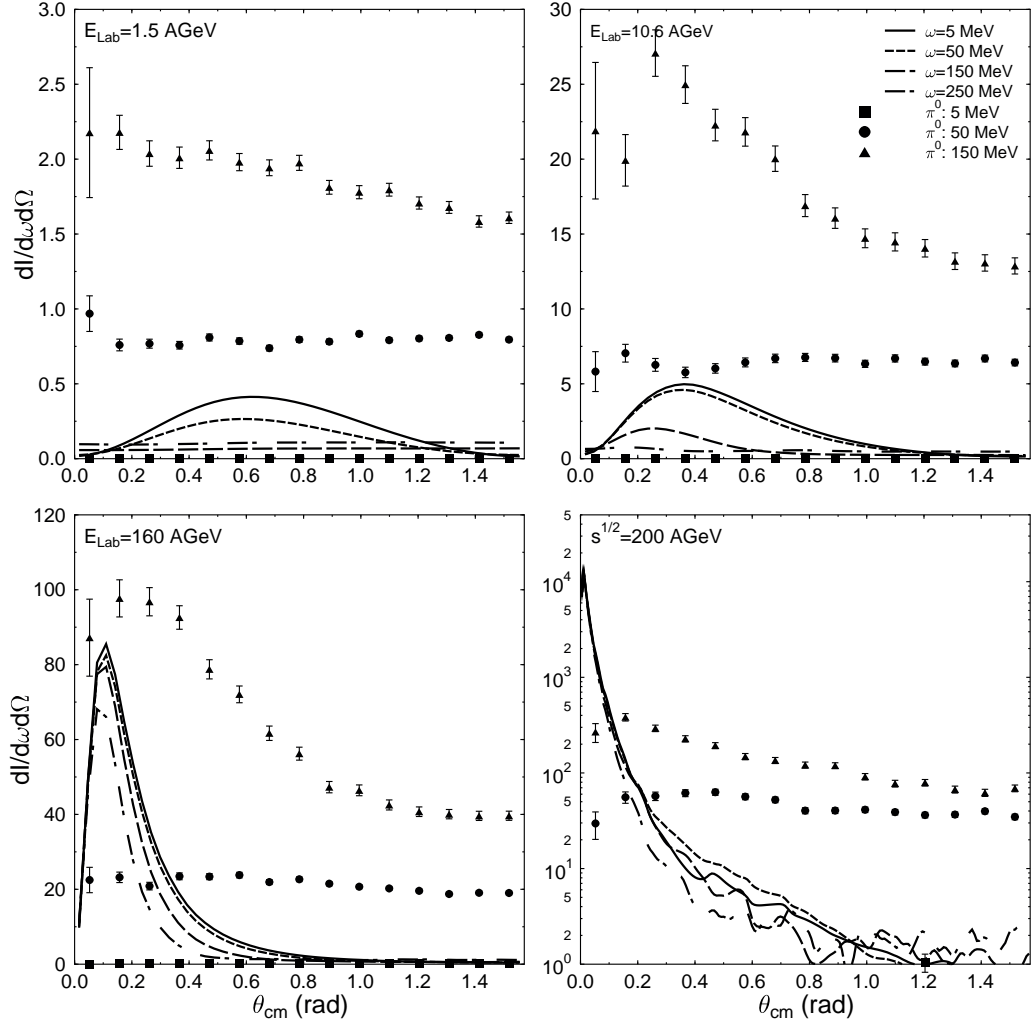,width=15cm}}
\caption{\label{all}Bremsstrahlung and $\pi^0$ decay spectra predicted by 
the UrQMD model for central Au+Au collisions 
at the SIS, AGS, SPS and RHIC energies.}
\end{figure}

This is illustrated in Figure~\ref{all} which 
shows the photon spectra predicted by the URQMD model at different
bombarding energies. In accordance with the general discussion the photon
spectra become more forward peaked with increasing initial energy. 
The background of electromagnetic decays of $\pi^0$
mesons 
at the AGS energy 
exceeds the coherent photon yield already at $\omega \gsim 50$~MeV. 
At the SPS and RHIC energies the coherent bremsstrahlung remains important
even for higher $\omega$'s (at small emission angles).

\section{Conclusions}
\label{sect_5}
We have discussed general properties of the
coherent bremsstrahlung spectra common to any model
of heavy-ion collisions. The comparison of
four different models shows that the photon spectra
are sensitive to the dynamics of the projectile-target
stopping only in the non--soft regime. 
We have shown, however, that even at higher photon energies 
different models may generate similar photon spectra.
Photon bremsstrahlung
produced in the stopping phase of a heavy-ion collision yields 
a dominant contribution to the soft photon spectra. 
As shown 
within the microscopic model the 
radiation 
of non-soft photons comes mainly 
from 
later (expansion) stages. The interference of this radiation 
with the emission from
the initial stage is relatively small.

The soft photon yield can provide some 
information about the degree of stopping. 
Soft photon spectra at ultrarelativistic bombarding energies, however, 
reveal
a rather weak sensitivity to the final velocities of secondary particles.

At the AGS and lower bombarding energies 
the 
background of $\pi^0$ decays 
is relatively large even for soft photons. 
According to our analysis, measurements of the soft photon spectra  
can hardly provide 
precise information about characteristics of nuclear stopping.

It is shown by general analysis that 
oscillations of the energy spectra 
always appear in simple models assuming a finite time of the collision and,
therefore, 
may be an artifact
of a simplified description. 
On the other hand, using 
these oscillations 
to extract properties of matter in nuclear collisions 
is only possible within
certain simplified models \cite{Bertulani}. 
Moreover, these oscillations are strongly smoothened if one takes into
account the photon radiation from the late expansion stages. 

The analysis of microscopic calculations performed within the UrQMD model 
shows that the shapes of the
bremsstrahlung 
spectra as functions of the photon energy are strongly 
sensitive to the late
stages of the reaction. In particular, 
this considerably complicates using 
spectral slopes as a 
tool to estimate the compression of matter in a heavy-ion
collision.

\section*{Acknowledgements}
This work was supported by
the Graduiertenkolleg Theoretische und Experimentelle
Schwerionenphysik, GSI, BMBF and DFG.

\newpage

\end{document}